\documentclass[12pt,preprint]{aastex}
\usepackage{txfonts}
\usepackage{amssymb}
\usepackage{lscape}

\shorttitle{A model for HST-1 in M87} \shortauthors{Liu et al.}

\begin{document}


\title{AN HOURGLASS MODEL FOR THE FLARE OF HST-1 IN M87}


\author{Wen-Po ~Liu\altaffilmark{1,2}, Guang-Yao ~Zhao\altaffilmark{2,3},
Yong Jun ~Chen\altaffilmark{2,4}, Chun-Cheng Wang\altaffilmark{5}, and Zhi-Qiang ~Shen\altaffilmark{2,4}}


\altaffiltext{1}{Correspondence author; College of Science, Civil Aviation University of
China, Tianjin 300300, China; email: wp-liu@cauc.edu.cn}
\altaffiltext{2}{Key Laboratory for Research in Galaxies and
Cosmology, Shanghai Astronomical Observatory, Chinese Academy of
Sciences, Shanghai 200030, China}
\altaffiltext{3}{Korea Astronomy and Space Science Institute, Daejeon 305-348, Korea}
\altaffiltext{4}{Key Laboratory of Radio Astronomy, Chinese Academy of Sciences, Nanjing, China}
\altaffiltext{5}{Key Laboratory for Research in Galaxies and Cosmology, University of Science and Technology of China, Hefei, Anhui 230026, China}


\begin{abstract}
To explain the multi-wavelength light curves (from radio to X-ray) of HST-1 in the M87 jet, we propose an hourglass model that is a modified two-zone system of Tavecchio \& Ghisellini (hereafter TG08)$\colon$ a slow hourglass-shaped or Laval nozzle-shaped layer connected by two revolving exponential surfaces surrounding a fast spine, through which plasma blobs flow. Based on the conservation of magnetic flux, the magnetic field changes along the axis of the hourglass. We adopt the result of TG08---the high-energy emission from GeV to TeV can be produced through inverse Compton by the two-zone system, and the photons from radio to X-ray are mainly radiated by the fast inner zone system. Here, we only discuss the light curves of the fast inner blob from radio to X-ray. When a compressible blob travels down the axis of the first bulb in the hourglass, because of magnetic flux conservation, its cross section experiences an adiabatic compression process, which results in particle acceleration and the brightening of HST-1. When the blob moves into the second bulb of the hourglass, because of magnetic flux conservation, the dimming of the knot occurs along with an adiabatic expansion of its cross section. A similar broken exponential function could fit the TeV peaks in M87, which may imply a correlation between the TeV flares of M87 and the light curves from radio to X-ray in HST-1. The Very Large Array (VLA) 22 GHz radio light curve of HST-1 verifies our prediction based on the model fit to the main peak of the VLA 15 GHz radio light curve.

\end{abstract}



\section{INTRODUCTION AND OBSERVATION CONSTRAINTS}
As is well known, HST-1 is the innermost knot of the M87 jet, located $\sim$80 pc from the core (Biretta et al. 1999).
The multi-wavelength light curves of HST-1 have been previously studied using radio data (Chang et al. 2010), optical and UV data (Perlman et al. 2003; Madrid 2009) and X-ray data (Harris et al. 2003, 2006, 2009). Chen et al. (2011) investigated the
radio polarization and spectral variability of HST-1, and Perlman et al. (2011) researched the optical polarization and spectral variability of the M87 jet. Cheung et al. (2007) argued that HST-1 may be the site of the flaring TeV gamma-ray emission reported by the H.E.S.S. (Aharonian et al. 2006). A study of particular note was by Abramowski et al. (2012), who
released 10 yr of multi-wavelength observations of M87 and the very high energy $\gamma$-ray flare of 2010. The quasi-simultaneous spectrum (from the radio to X-ray band; e.g., Marshall et al. 2002; Waters \& Zepf 2005; Perlman \& Wilson 2005; Harris et al. 2006; Cheung et al. 2007) and polarization observations (e.g., Perlman et al. 1999, 2011; Chen et al. 2011) of the M87 knots demonstrate the nature of synchrotron radiation.

From the multi-wavelength light curves of HST-1 (Fig. 1), we could obtain some physical constrains on HST-1: the light curves show two big flares, with the main peaks of the light curves around the year 2005.30 and the second ones around the year 2007.

Here, we discuss a pure (or single) process (Doppler effect). For synchrotron emission in the case of a moving sphere, with observed fluxes $I_{\nu,\mbox{\scriptsize obs}}=\delta^{3+\alpha}I_{\nu}$ (Dermer 1995; $\delta$ is the Doppler factor of HST-1 and $\alpha=(p-1)/2$, where $p$ is the spectral index of the particles), the change of the Doppler factor may explain the change in the light curve. However, Harris et al. (2006) suggested a modest beaming synchrotron model with a Doppler factor of three or four, while Wang \& Zhou (2009) obtained the Doppler factor of HST-1 to be $3.57\pm0.51$ through a synchrotron model fitting. The HST-1 complex could be model fitted with multiple components (e.g., Cheung et al. 2007; Giroletti et al. 2012); in this case, the speed of one component in HST-1 is obtained through a longer monitoring of the same one, but the velocity estimate usually has fairly large uncertainties. Giroletti et al. (2012) reported the apparent velocities of very long baseline interferometry (1.7 GHz Very Long Baseline Array, VLBA and 5 GHz European VLBI Network, EVN) components in HST-1 with high precision ($<$$2\%$, this is an unprecedented accuracy for determining the apparent velocities of components in HST-1) during the decay period of the HST-1 flare, which implies that components in HST-1 have uniform motion with high precision. Hence, the variation range of the Doppler factor may be very small and the change in the Doppler factor in HST-1 may not explain the order of the flux change.

Excluding the aforementioned single mechanism, we believe that the flare of HST-1 may be correlated with some complex processes which include a changing magnetic field strength.

In Section 2, we describe in detail our model for HST-1. In Section 3, we present and discuss the fitting results of this model to the main peak in the multi-wavelength light curves of HST-1. A summary is provided in Section 4.

\section{The Model}
Tavecchio \& Ghisellini (2008, TG08) suggested a two-zone scenario in subparsec-scale jets to explain the TeV emissions in M87: a slow hollow cylindrical layer (the velocity relative to the M87 core is $v_l$) surrounds a fast cylindrical zone (the velocity relative to the M87 core is $v_s$, $v_s\gg v_l$, and the velocity of the inner zone relative to the cylindrical layer is $v$). The high energy from GeV to TeV could be produced through inverse Compton by the two-zone, and the photons from radio to X-ray are mainly radiated by the fast inner zone. If this subparsec-scale structure is located within HST-1, HST-1 may be a TeV emission source. However, the model of TG08 could not explain the multi-wavelength light curves (or flare) from radio to TeV. We found that a modified scheme of the TG08 model could achieve this; here we only discuss the light curves of the inner blob from radio to X-ray. We believe that the slow layer may be an hourglass-shaped or Laval nozzle-shaped layer connected by two revolving exponential surfaces (Fig. 2). Considering magnetic field conservation, we believe that the magnetic field will change along the axis of the hourglass. We assumed that the length of the inner blob along the jet axis may be smaller than the radius $R_s$ of its cross section, which may be subparsec-scale, and may remain unchanged (the blob may be constrained in a series of shocks along the jet axis). If the axis coordinate and radius of the hourglass nozzle are $x_n$ and $R_n$, respectively, then the layer radius $R$ (we assumed that $R\sim R_s$) as a broken exponential function of the axis coordinate is
\begin{equation}
R=\left\{
\begin{array}{ll}
R_n e^{k(x_n-x)},&\mbox{~$x<x_n$;}\\
R_n e^{k^{'}(x-x_n)},&\mbox{~$x>x_n$,}\\
\end{array}
\right. \end{equation}

\noindent where $k$ and $k^{'}$ are constant.

Considering magnetic field conservation, $B\propto R^{-2}$, the magnetic field $B$ along the axis of the hourglass will be
\begin{equation}
B=\left\{
\begin{array}{ll}
B_n e^{-2k(x_n-x)},&\mbox{~$x<x_n$;}\\
B_n e^{-2k^{'}(x-x_n)},&\mbox{~$x>x_n$,}\\
\end{array}
\right. \end{equation}
\noindent where $B_n$ is the magnetic field of the hourglass nozzle.

When a blob travels down the axis of the first bulb in the hourglass, because of magnetic field conservation, its cross section experiences an adiabatic compression (compression velocity $u=-(dR/dt)=R_n k v e^{k(x_n-x)}$), which results in particle acceleration and the brightening of HST-1 (this may explain why Perlman et al. 2011 found no evidence for the motion of the flaring blob of HST-1 in their optical data from \emph{Hubble Space Telescope} (\emph{HST}) observations but Giroletti et al. 2012 found no prominent stationary components in their radio data from VLBA and EVN observations). When the blob moves into the second bulb of the hourglass, considering magnetic field conservation, the dimming of the knot will occur along with an adiabatic expansion of its cross section ($u=-(dR/dt)=-R_n k^{'} v e^{k^{'}(x-x_n)}$).

When a blob approaches the hourglass nozzle, the energy gains of the particles in the blob are $(\partial E/\partial t)=\alpha_2 E$, where $\alpha_2=(2/3)(u/R)=(2/3)kv>0$ (Pacholczyk 1970). The compression timescale $\tau=(R/u)=(2/3\alpha_2)$.

A modified continuous injection model, which considers that source particles (a broken power law which could be interpreted as partial loss or escape of high-energy particles in the acceleration region) are continuously injected into an adjacent radiation region from an acceleration region, can be fit to the spectral energy distribution of the outer jet in M87 (Liu \& Shen 2007; Sahayanathan 2008 presented a similar model with two spectral indices). We suppose that the radiation mechanism of the inner jet is the same as that of the outer jet. When the source particles (a broken power law) have been injected for the time interval of $t_1$, the radiation region of a blob in HST-1 moves into the hourglass layer. Then, the initial condition of the particle spectra (this is a relic of a past process and may not be ignored for the next process) in HST-1 at the beginning of the compression process is (Liu \& Shen 2007; Sahayanathan 2008)

\begin{equation}
N^{*}(E, \theta, 0)=\left\{
\begin{array}{ll}
q_1^{*} t_1 E^{-p_1},&\mbox{~$E\ll E_b$;}\\
q_2^{*} t_1 E^{-(p_2+1)},&\mbox{~$E_b\ll E\ll \frac{1}{\beta_0 t_1}$,}\\
\end{array}
\right. \end{equation}

\noindent where $N^{*}$, $q_1^{*}$, and $q_2^{*}$ refer to the entire radiation region in the blob taken as an entity (but $N$, $q_1$, and $q_2$ are applied to a unit volume); $\theta$ is the pitch angle between the magnetic field and the particle; $E_b$ is the break energy of a broken power law (here, we assumed that $E_b<(1/\beta_0 t_1)$; Sahayanathan 2008); $p_1$ and $p_2$ denote particle spectrum indices that may be different (e.g., Sahayanathan 2008); $\beta_0=bB^2_{\perp}$; $b$ is a constant; and $B_{\perp}$ represents the component of the magnetic field perpendicular to the velocity of the particle.

Because the acting timescale of the compression process is shorter than that of the particle injection process, which may be comparable with the kinetic timescale of HST-1 relative to the M87 core, we can ignore the influence of particle injection in the compression process. Next, we consider synchrotron radiation and an adiabatic compression of the radiation region in the blob, and the kinetic equation is (Kardashev 1962)

\begin{equation}
\frac{\partial N^{*}}{\partial t}=-\alpha_2 \frac{\partial}{\partial
E}(EN^{*})+\beta \frac{\partial}{\partial E}(E^2 N^{*}),
\end{equation}

Under the initial conditions given in Equation (3), we have

\begin{equation}
N^{*}(E, \theta, t)=
q^{*}t_1 E^{-\lambda }[1-E e^{-a_2}\int_{0}^{t} \beta e^{a_2} dt]^{\lambda-2}e^{(\lambda-1)a_2},
\end{equation}

\noindent where $a_2=\int_{0}^{t} \alpha_2 dt$, $q^{*}$ represents $q_1^{*}$ or $q_2^{*}$, $\lambda$ is a substitute for $p_1$ or $p_2+1$, and we have assumed that $E\ll (e^{a_2}/\int_{0}^{t} \beta e^{a_2} dt)$ and $\alpha_2=$const in the formula (5). Hence,

\begin{equation}
N^{*}(E, \theta, t)=\left\{
\begin{array}{ll}
q_1^{*} t_1 E^{-p_1}e^{(p_1-1)\alpha_2 t}[1-\frac{\beta}{7\alpha_2  }E(1-e^{-7\alpha_2 t})]^{p_1-2},&\mbox{~$E\ll E_b$;}\\
q_2^{*} t_1 E^{-(p_2+1)}e^{p_2\alpha_2 t}[1-\frac{\beta}{7\alpha_2}E(1-e^{-7\alpha_2 t})]^{p_2-1},&\mbox{~$E_b\ll E\ll \frac{7\alpha_2}{\beta (1-e^{-7\alpha_2 t})}$.}\\
\end{array}
\right. \end{equation}

Considering the conservation of magnetic flux, $B\propto R^{-2}$, $R=R_0 e^{-k v t}=R_0 e^{-(3/2)\alpha_2 t}$, where $R_0$ is the initial radius of the blob. Hence, $\beta=\beta_0 e^{6\alpha_2 t}$. The factors $[1-(\beta/7\alpha_2)E(1-e^{-7\alpha_2 t})]^{p_1-2}$ and $[1-(\beta/7\alpha_2)E(1-e^{-7\alpha_2 t})]^{p_2-1}$ in Formula (6) are close to 1, because of the following reasons. (1) $E\ll (7\alpha_2/\beta (1-e^{-7\alpha_2 t}))$, and so $1-(\beta/7\alpha_2)E(1-e^{-7\alpha_2 t})\rightarrow 1$. (2) $p_1, p_2\sim 2$ in the M87 jet (Perlman \& Wilson
2005; Liu \& Shen 2007; Perlman et al. 2011), hence $p_1-2\sim 0$ and $p_2-1\sim 1$.

Formula (6) is thus close to
\begin{equation}
N^{*}(E, \theta, t)\approx \left\{
\begin{array}{ll}
q_1^{*} t_1 E^{-p_1}e^{(p_1-1)\alpha_2 t},&\mbox{~$E\ll E_b$;}\\
q_2^{*} t_1  E^{-(p_2+1)}e^{p_2\alpha_2 t},&\mbox{~$E_b\ll E\ll \frac{7\alpha_2}{\beta (1-e^{-7\alpha_2 t})}$,}\\
\end{array}
\right. \end{equation}

and when this is converted to unit volume (entire volume $V\propto R^2 \propto e^{-3\alpha_2 t}$),
\begin{equation}
N(E, \theta, t)=\frac{N^{*}}{V}\approx \left\{
\begin{array}{ll}
q_1 t_1 E^{-p_1}e^{(p_1+2)\alpha_2 t},&\mbox{~$E\ll E_b$;}\\
q_2 t_1 E^{-(p_2+1)}e^{(p_2+3)\alpha_2 t},&\mbox{~$E_b\ll E\ll \frac{7\alpha_2}{\beta (1-e^{-7\alpha_2 t})}$.}\\
\end{array}
\right. \end{equation}

Next, we consider the increase in magnetic field strength and the time dependence of blob
length along the line of sight ($\propto e^{-(3/2)\alpha_2 t}$). If the distributions
of particles are isotropic, we can derive the flux formula of the synchrotron model:
\begin{equation}
I_{\nu}\propto\left\{
\begin{array}{ll}
e^{(5p_1+4)\alpha_2   t/2}\nu^{-(p_1-1)/2},&\mbox{~$\nu\ll \nu_{B1}$;}\\
e^{(5p_2+9)\alpha_2   t/2}\nu^{-p_2/2},&\mbox{~$\nu_{B1}\ll \nu\ll \nu_{B2}$,}\\
\end{array}
\right. \end{equation}
\noindent where $\nu_{B1}$ and $\nu_{B2}$ are break frequencies.

When the radiation region of the blob in HST-1 moves away from the hourglass and into the adjacent bulb, deceleration of particles in the blob may occur. Then, $\alpha_2 ^{'}=-(2/3)k^{'}v<0$, the expansion timescale $\tau^{'}=-(2/3\alpha_2 ^{'})$, and the formula for the particle spectrum will become
\begin{equation}
N^{'}(E, \theta, t)\approx \left\{
\begin{array}{ll}
q_1 t_1 E^{-p_1}e^{(p_1+2)\alpha_2  t_2}e^{(p_1+2)\alpha_2 ^{'} (t-t_2)},&\mbox{~$E\ll E_b$;}\\
q_2 t_1 E^{-(p_2+1)}e^{(p_2+3)\alpha_2  t_2}e^{(p_2+3)\alpha_2 ^{'} (t-t_2)},&\mbox{~$E_b\ll E\ll \frac{7\alpha_2^{'}}{\beta^{'} [1-e^{-7\alpha_2^{'}} (t-t_2)]}$,}\\
\end{array}
\right. \end{equation}

\noindent where $t_2$ is the acting timescale of the compression process, $\beta^{'}\propto e^{6 \alpha_2^{'} (t-t_2)}$.

Now, we consider the decrease of magnetic field strength and the time dependence of blob length along the line of sight ($\propto e^{-(3/2)\alpha_2^{'} (t-t_2)}$). With this consideration, the flux expression would be changed to
\begin{equation}
I_{\nu}^{'}\propto\left\{
\begin{array}{ll}
e^{(5p_1+4)\alpha_2 ^{'}  t/2}\nu^{-(p_1-1)/2},&\mbox{~$\nu\ll \nu_{B1}$;}\\
e^{(5p_2+9)\alpha_2 ^{'}  t/2}\nu^{-p_2/2},&\mbox{~$\nu_{B1}\ll \nu\ll \nu_{B2}^{'}$,}\\
\end{array}
\right. \end{equation}

\noindent where $\nu_{B1}$ and $\nu_{B2}^{'}$ are break frequencies. According to Formulae (9) and (11), the flux $f$ changes
exponentially with time in our model, i.e., for low
frequency,
\begin{equation}
f\propto\left\{
\begin{array}{ll}
e^{(5p_1+4) \alpha_2 t/2},&\mbox{~$t<t_{\mbox{\scriptsize peak}}$;}\\
e^{(5p_1+4) \alpha_2 ^{'} t/2},&\mbox{~$t>t_{\mbox{\scriptsize peak}}$,}\\
\end{array}
\right. \end{equation}

\noindent where $t_{\mbox{\scriptsize peak}}$ denotes the peak time of the light curve.

Further, for high frequency,
\begin{equation}
f\propto\left\{
\begin{array}{ll}
e^{(5p_2+9) \alpha_2 t/2},&\mbox{~$t<t_{\mbox{\scriptsize peak}}$;}\\
e^{(5p_2+9)\alpha_2 ^{'} t/2},&\mbox{~$t>t_{\mbox{\scriptsize peak}}$.}\\
\end{array}
\right. \end{equation}

\section{FITTING RESULTS AND DISCUSSION}
Now, we apply the above hourglass model to the multi-wavelength light curves of HST-1 in the M87 jet. Here, the beaming factor
of a blob is assumed to be constant.

The data we used are plotted in Fig. 1. The units of the radio data are 1 Jy. VLBA 15 GHz radio data from Chang et al. (2010), in which relative uncertainty is assumed to be $5\%$, are plotted as up triangles and down triangles that show the
upper limits of fluxes. Very Large Array (VLA) 15 GHz radio data from Harris et al. (2009) and Abramowski et al. (2012) are plotted as squares. The units of the UV data are 1 mJy (Madrid 2009); these data are plotted as circles. For the X-ray data, we use the flux density integrated from 0.2 to 6 keV (Harris et al. 2006, 2009), and the units are
$10^{-11}$ erg cm$^{-2}$ s$^{-1}$; these data are plotted as diamonds. The X-ray data after 2005 August 6 were estimated by assuming that the correction factor for the (unknown) effective area (Harris et al. 2006) is the same as that for 2005 August 6. Harris et al. (2006) estimate that the resulting uncertainties are of the order of $15\%$. Note that because of the unknown effective area, the uncertainties of the X-ray fluxes are larger than those in the counts, which may reduce (or smooth) the individual oscillations of the X-ray fluxes. In our model, as the time dependence of the effective area is considered, the resulting uncertainties of the X-ray fluxes may be larger than $15\%$.

Based on the shape trends of the radio, UV, and X-ray light curves in HST-1, as shown in Fig. 1, we select a common shape section (from about the year 2003 to 2006.60), which may be the main peak in our fitting area. We used the weighted least-squares method to fit our model to this main peak, with Equation (12) corresponding to the radio and UV light curves and
Equation (13) corresponding to the X-ray light curve. Although there is a peak time in our model, its exact position was unknown in the fitting of our model to the light curve. Hence, we first arbitrarily divided the light curve data during the chosen period into two groups to perform the least-squares method, in which the sum of the reduced chi square $\chi^2_{\nu}$ for the two parts is the least. Then we can calculate the corresponding least $\chi^2_{\nu}$ by changing the division of the two groups. All reasonable combinations of the two groups (e.g., the fitting peak time should lie between the two groups) are considered before we obtain the best fit with a minimal $\chi^2_{\nu}$. These best-fit parameters for each light curve are shown in Fig. 1 by blue solid lines. Note that the reduced chi square value depends on uncertainties of the data.

Our model can well fit the main peak in the multi-wavelength light curves of HST-1 as shown in Fig. 1. It satisfies the aforementioned first constraint for the multi-wavelength light curves of HST-1: the main peak time is around the year 2005.30, as shown in Fig. 1 and Table 1. The stratified effect of radiation regions in the outer knots of M87 has been verified by Perlman et al. (1999), Marshall et al. (2002), and Perlman $\&$ Wilson (2005). The flattened peak section in the radio light curve can be explained by the greater length of the radio radiation region along the jet axis than the UV and X-ray region. Abramowski et al. (2012) showed that the VLA 22 GHz radio light curve of HST-1 is consistent with the 15 GHz light curve within the error range, which implies that our model can also explain the VLA 22 GHz light curve. In other words, the observation verifies our prediction based on the model fit to the main peak of the VLA 15 GHz light curve.

The aforementioned case is a single component; however, a complex structure that contains multiple components may be more reasonable, as in this case, each peak corresponds to a component passing through the hourglass nozzle. We also use the aforementioned method to fit our model to the possible secondary peak of the multi-wavelength light curves. The best fits for our model are plotted in Fig. 1 in red dotted and dot-dashed lines.

A similar broken exponential function could fit the TeV peaks in M87 (Abramowski et al. 2012), which may imply a correlation between the TeV flares of M87 and the light curves from radio to X-ray in HST-1. Further, the maximum of the TeV flares of M87 was coincident with the peak of light curves from radio to X-ray in HST-1 observed in 2005 (Cheung et al. 2007); this may imply that the observable TeV flux density was produced through inverse Compton as a blob passed through the hourglass nozzle. The detailed generation process of the TeV flare may be very complicated.

The estimates of $p_1$ and $p_2$ in HST-1 require simultaneous broad observational data, which are scarce. Liu \& Shen (2007) found that the averaged spectral index of outer knots in M87 is about 2.36 through model fitting. If we assumed that the spectral index of HST-1 is similar to that of outer knots (i.e., $p_1, p_2\approx 2.36$), we could derive some physical parameters of HST-1 from our model fits, as shown in Table 2. Based on the timescale of $\sim$ 5.6 yr, the local size of the component is smaller than a parsec. The derived $\alpha_2$ agrees well with the derived $\alpha_2^{'}$, and the common value is around 0.12, which implies that the hourglass-shaped layer may be symmetrical with respect to the nozzle.

\section{CONCLUSION}
We propose a modified two-zone system of TG08: a fast blob passes through a slow hourglass-shaped layer that is connected by two revolving exponential surfaces. Mainly, the emissions from radio to X-ray are radiated by the fast blob. Because of magnetic flux conservation, the brightening and dimming of HST-1 could be explained as adiabatic compression and expansion, respectively, of a blob passing through the outer layer. The observable TeV flux density may be produced through inverse Compton as a blob passes through the nozzle. The VLA 22 GHz radio light curve of HST-1 was used to verify our model.

\section{ACKNOWLEDGMENT}
We are grateful to Prof. D. E. Harris for his help in the VLA 15 GHz radio light curve of HST-1. We acknowledge the support from the National Natural Science Foundation of China (NSFC) through grants U1231106 and 11273042, the Science and Technology Commission of Shanghai Municipality (12ZR1436100), and the Scientific Research Foundation of the Civil Aviation University of China (09QD15X). This work is partly supported by the China Ministry of Science and Technology under the State Key Development Program for Basic Research (2012CB821800), NSFC (grants 10625314, 11121062, 11173046, 11033007, 10973012, and 11073019), the CAS/SAFEA International Partnership Program for Creative Research Teams, the Strategic Priority Research Program on Space Science, the Chinese Academy of Sciences (Grant No. XDA04060700), and 973 program 2007CB815405.

\clearpage
\begin{landscape}
\setlength{\headsep}{1cm}
\begin{deluxetable}{crrrrrrrr}
\tabletypesize{\small} \tablewidth{0pc} \tablecaption{Parameters for Model Fits to the Main peak of the Radio, UV and X-Ray Light Curves \label{tab:modelX}}
\tablehead{\colhead{Wave Band}  & \colhead{$(5p_1+4)\alpha_2/2$} &
\colhead{$(5p_2+9)\alpha_2/2$} & \colhead{$(5p_1+4){\alpha_2}^{'}/2$} &
\colhead{$(5p_2+9) {\alpha_2}^{'}/2$} & \colhead{Peak Time (yr)} &
\colhead{$\chi^2_{\nu}$ (dof)}} \startdata

  X-ray & $\cdots$ & $1.33\pm0.09$ & $\cdots$ & $-1.33\pm0.07$ & $2005.26^{+0.04}_{-0.14}$ & $1.35$ (26)\\
  UV & $0.88\pm0.04$ & $\cdots$ & $-0.85\pm0.12$ & $\cdots$ & $2005.37^{+0.10}_{-0.02}$ & $8.99$ (17)\\
  Radio (VLA 15 GHz) & $0.80\pm0.06$ & $\cdots$ & $-1.14\pm0.17$ & $\cdots$ & $2005.67^{+0.46}_{-0.34}$ & $0.66$ (5)\\
  Radio (VLBA 15 GHz) & $1.02\pm0.13$ & $\cdots$ & $-0.92\pm0.51$ & $\cdots$ & $2005.04^{+0.26}_{-0.12}$ & $40.80$ (2)\\
  \enddata
\tablecomments{Column 7: reduced chi square for the number of degrees of freedom given in parentheses. Note that the reduced chi square value depends on data uncertainties and the main peak chosen is about year 2003 -- 2006.60.}
\end{deluxetable}
\end{landscape}

\begin{landscape}
\setlength{\headsep}{1cm}
\begin{deluxetable}{crrrrrrrrrr}
\tabletypesize{\small} \tablewidth{0pc} \tablecaption{The Parameters Derived from Model Fits}
\tablehead{\colhead{Parameters} & \colhead{$\alpha_2$} & \colhead{$\tau$ (yr)} & \colhead{$\alpha_2^{'}$} & \colhead{$\tau^{'}$ (yr)}} \startdata
  X-ray & $0.13\pm 0.01$ & $5.21\pm 0.35$ & $-0.13\pm0.01$ & $5.21\pm 0.27$ \\
  UV & $0.11\pm 0.01$ & $5.98\pm 0.27$ & $-0.11\pm0.02$ & $6.20\pm 0.87$ \\
  Radio (VLA 15 GHz) & $0.10\pm 0.01$ & $6.58\pm 0.49$ & $-0.14\pm0.02$ & $4.62\pm 0.69$ \\
  Radio (VLBA 15 GHz) & $0.13\pm 0.02$ & $5.16\pm 0.66$ & $-0.12\pm0.06$ & $5.72\pm 3.17$ \\
\enddata
\tablecomments{The parameters ($\alpha_2$, $\alpha_2^{'}$, the compression timescale $\tau$ and the expansion timescale $\tau^{'}$) derived from Table 1 with $p_1=p_2=2.36$. }
\end{deluxetable}
\end{landscape}

\clearpage
\begin{figure}
\includegraphics[]{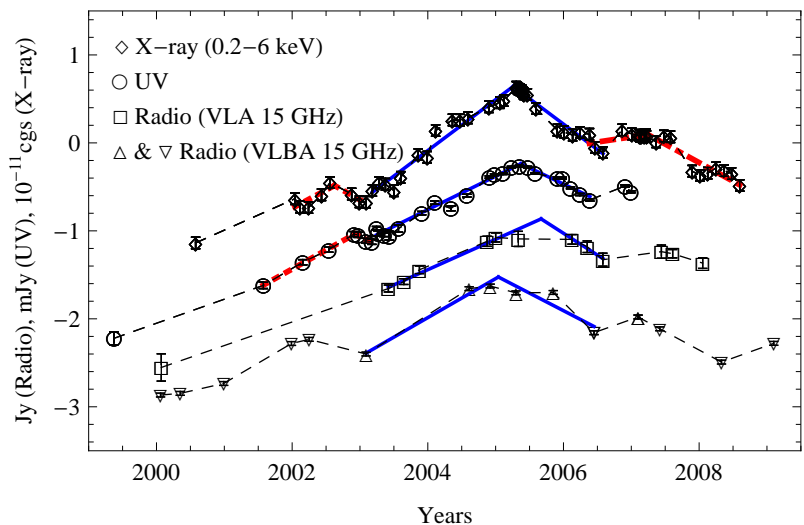}
\caption{Multi-wavelength light curves for HST-1. The plot on a log scale shows the fluxes of radio, UV, and X-ray. The units of the radio data are 1 Jy. VLBA 15 GHz radio data from Chang et al. (2010), are plotted as up triangles and down triangles that show the upper limits of the fluxes. VLA 15 GHz radio data from Harris et al. (2009) and Abramowski et al. (2012) are plotted as squares. The units of the UV data are 1 mJy (Madrid 2009); these data are plotted as circles. For the X-ray data, we use the flux density integrated from 0.2 to 6 keV (Harris et al. 2006, 2009), and the units are $10^{-11}$ erg cm$^{-2}$ s$^{-1}$; these data are plotted as diamonds. The black dashed lines directly join the data. The blue solid lines show the fits for our model to the selective data from about year 2003 to 2006.60. The best fits for our model to some possible peaks of the multi-wavelength light curves are plotted in red dotted and dot-dashed lines.}
\end{figure}

\clearpage
\begin{figure}
\includegraphics[]{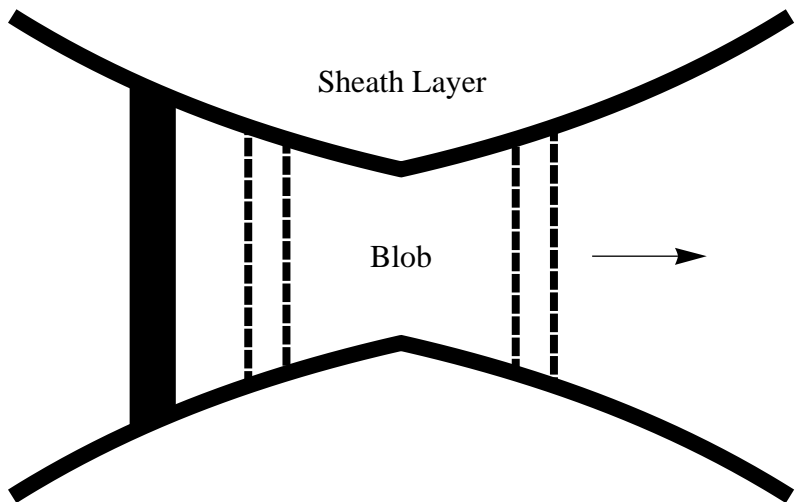}
\caption{Schematic illustrating the axis section of an hourglass model that is a modified two-zone system of TG08. A compressible blob passes through an hourglass-shaped or Laval nozzle-shaped sheath layer that is connected by two revolving exponential surfaces.}
\end{figure}

\end{document}